\documentclass[prl,10pt,superscriptaddress,twocolumn]{revtex4}

\usepackage{graphicx}
\usepackage{amsmath}
\usepackage{color}
\usepackage{units}
\usepackage[latin1]{inputenc}
\usepackage{verbatim}
\usepackage{float}
\usepackage[english]{babel}
\usepackage{booktabs}	
\usepackage{epstopdf}
\usepackage{pgf}
\usepackage{pgfplots}
\usepackage{placeins}

\begin{document}


\title{Reduction of Classical Measurement Noise via Quantum-Dense Metrology}

\author{Melanie\,Ast}
\affiliation{Institut f\"ur Laserphysik und Zentrum f\"ur Optische Quantentechnologien der Universit\"at Hamburg,\\%
Luruper Chaussee 149, 22761 Hamburg, Germany}
\affiliation{Institut f\"ur Gravitationsphysik der Leibniz Universit\"at Hannover and\\ %
Max-Planck-Institut f\"ur Gravitationsphysik (Albert-Einstein-Institut), Callinstra{\ss}e 38, 30167 Hannover, Germany}%

\author{Sebastian\,Steinlechner}
\affiliation{Institut f\"ur Gravitationsphysik der Leibniz Universit\"at Hannover and\\ %
Max-Planck-Institut f\"ur Gravitationsphysik (Albert-Einstein-Institut), Callinstra{\ss}e 38, 30167 Hannover, Germany}%
\affiliation{SUPA, School of Physics and Astronomy, University of Glasgow, Glasgow G12 8QQ,
United Kingdom}%

\author{Roman Schnabel}
\email[corresponding author:\\]{roman.schnabel@physnet.uni-hamburg.de}
\affiliation{Institut f\"ur Laserphysik und Zentrum f\"ur Optische Quantentechnologien der Universit\"at Hamburg,\\%
Luruper Chaussee 149, 22761 Hamburg, Germany}
\affiliation{Institut f\"ur Gravitationsphysik der Leibniz Universit\"at Hannover and\\ %
Max-Planck-Institut f\"ur Gravitationsphysik (Albert-Einstein-Institut), Callinstra{\ss}e 38, 30167 Hannover, Germany}%

\begin{abstract} 
Quantum-dense metrology (QDM) constitutes a special case of quantum metrology in which two orthogonal phase space projections of a signal are simultaneously sensed beyond the shot noise limit. Previously it was shown that the additional sensing channel that is provided by QDM contains information that can be used to identify and to discard corrupted segments from the measurement data. Here, we demonstrate a proof-of-principle experiment in which this information is used for improving the sensitivity without discarding any measurement segments. Our measurement reached sub-shot-noise performance although initially strong classical noise polluted the data.
\end{abstract}
\maketitle

\textbf{Introduction: } --
Measurement devices are usually designed and constructed in such a way that the measurement signal couples with a high signal-to-noise-ratio to a single readout observable. The orthogonal, non-commuting observable then usually does not contain any information about the signal, and is thus not read out. 
Let us consider a Michelson-type interferometer as it is currently used for the detection of gravitational waves \cite{AdvLIGO2015,AdvVirgo2015,GEOHF2014,GW150914}. Here, the signal is a change of the differential arm length, and the readout observable is the corresponding change of the output light's amplitude quadrature. The change of the output light's phase quadrature is not monitored since it does not contain any signal. The simultaneous measurement of non-commuting observables of a beam of light, on the first sight, is only disadvantageous since it requires splitting the output beam into two beams, which reduces the signal power and thus the signal-to-shot-noise ratio.
Recently, however, it was shown that the simultaneous measurement of the orthogonal amplitude and phase quadratures of the output light of an interferometer can result in an improved overall measurement sensitivity, if not quantum noise but excess noise that appears in both phase space projections limits the sensitivity \cite{ScM2015}. 
The most relevant kind of such excess noise is light that is first scattered out the main bright optical mode, hits surfaces having a relative motion, and is back-scattered into the main optical mode. Such `parasitic interferences' \cite{Vahlbruch2007} are indeed a problem in high-power laser interferometers for the detection of gravitational waves \cite{Fiori2010,Ottaway2012,AdvDetBook,Martynov2015,AdvLIGOSens2016}.

In quantum metrology \emph{nonclassical} states are used to read out the signal. A prominent example of quantum metrology is the application of squeezed light in the gravitational-wave (GW) detector GEO\,600 \cite{Schnabel2010,SqGEO2011,ltSqGEO2013}. An improved sensitivity due to squeezed light was also demonstrated in the LIGO detector at Hanford \cite{SqLIGO2013} and is earmarked as one of the next upgrades of the AdvancedLIGO detectors \cite{LIGOWhitePaper2015}. Future improvements of gravitational-wave detectors will also include a further increase of light powers in the arms, which further increases the risk of parasitic interferences. Therefore, it exists a strong motivation to investigate whether the simultaneous detection of two orthogonal phase space projections of a signal on a split output field can also be combined with quantum noise squeezing. 

In the case of squeezing enhanced metrology, however, equally splitting the output field into two beams not only reduces the signal power by a factor of two but also significantly reduces the squeezing factor. Furthermore, the quantum noise in the orthogonal observable is anti-squeezed, which reduces the amount of extractable information about disturbances.
Recently, quantum dense metrology (QDM) was proposed, which uses two-mode squeezing (entanglement with Gaussian quantum statistics) to circumvent the squeezing loss as well as the anti-squeezing issue \cite{SBM2013}. Without loosing squeezing and without degrading the additional channel with anti-squeezing, QDM was shown to provide additional information for identifying and discarding (`vetoing') corrupted segments from the measurement data. 

Here we demonstrate that the additional information in QDM offers the possibility for improving the sensitivity without discarding any measurement segments, by removing the classical disturbances from the channel that contains the signal. We demonstrate in a proof-of-principle experiment that the measurement sensitivity can be improved from above-shot-noise to sub-shot-noise (sub-Poissonian) performance. 
The way the excess noise arises does not have to be known but needs to be deterministic, allowing for fitting a model of the excess noise to the readout data.


\begin{figure}
 \center 
  \includegraphics[width=\columnwidth]{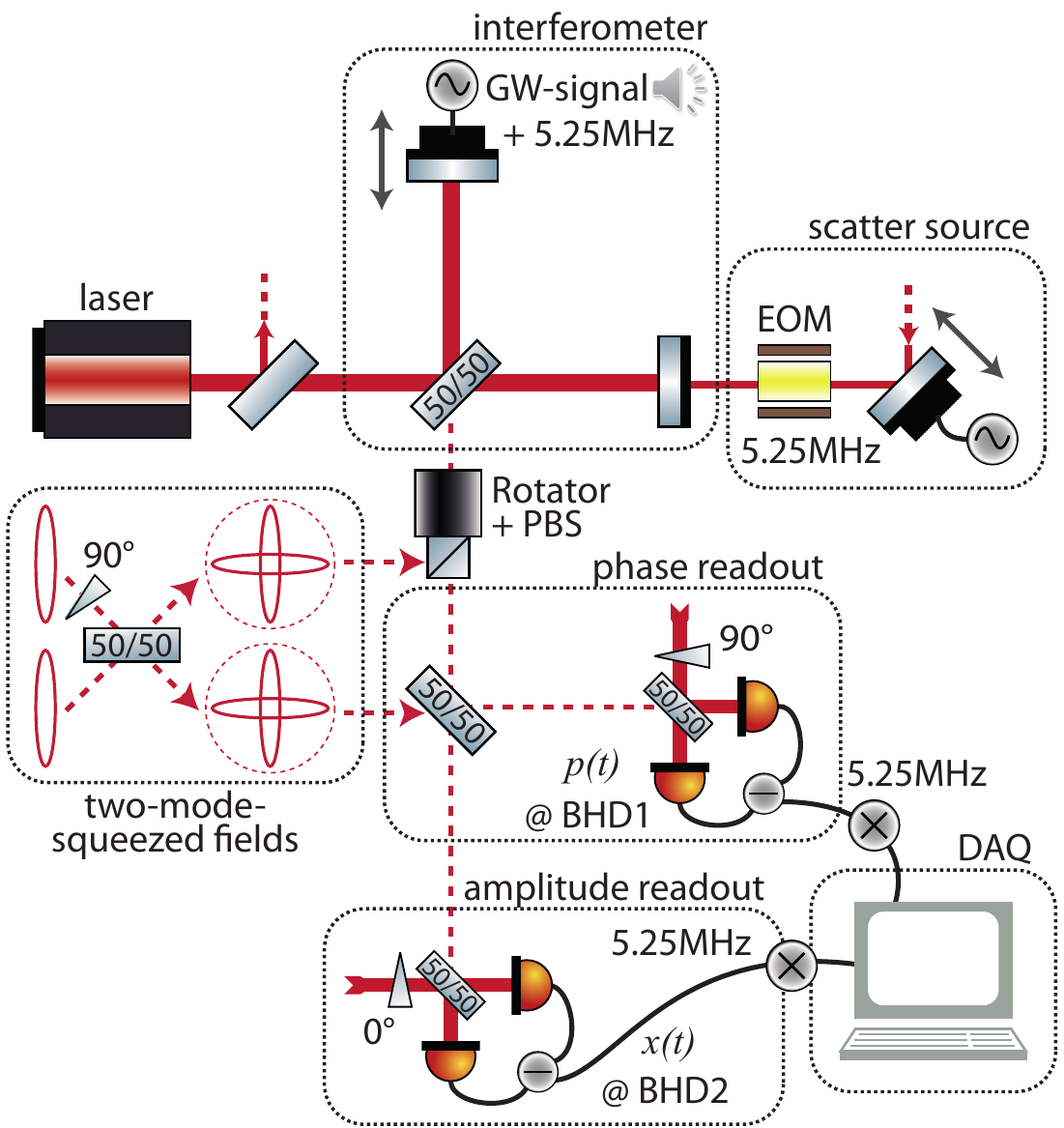}
  \caption{\textbf{Experimental setup.} In a table-top Michelson interferometer two signals were generated: a differential arm length change (a GW-like signal) by modulating a piezo mounted interferometer end mirror and one broadband parasitic interference using an external beam that we injected through the end mirror of the second arm. The output signal was split and detected with two balanced homodyne detectors (BHD1\&2), reading out the orthogonal amplitude and phase quadratures. The readout was enhanced with two-mode-squeezed states of light. One of their two subsystems was reflected at the interferometer's dark port via a combination of a Faraday rotator and a polarising beam splitter (PBS) while the other one was sent directly to the beam splitter in the interferometer output. DAQ: data acquisition system.}
\label{Fig1-QDM-Setup}
\end{figure}%

\textbf{Experimental setup: } -- 
Our table-top experiment combined two setups described in \cite{SBM2013, ScM2015}. A simplified schematic is shown in Fig.\,\ref{Fig1-QDM-Setup}. We employed a continuous-wave Nd:YAG laser source with an output power of 2\,W at wavelength $\lambda=1064\,\text{nm}$, which was split to supply the different parts of the experiment. In a simple Michelson interferometer of about 7\,cm arm length and with an input power of about 10\,mW we generated two test signals: one GW-like signal due to a differential arm length change and one scattered light disturbance coming from an external source. 
The GW-like signal we injected from a sound file containing about 4.5\,s of a simulated inspiral of two neutron stars with equal masses \cite{gwSound}. It started at about 55\,Hz and increased in frequency and amplitude over time. The signal was shifted in frequency by 5.25\,MHz with an Agilent 33500B series waveform generator and used to modulate the position of an interferometer end mirror that was mounted on a piezoelectric element.

The back-scatter disturbance we produced by injecting an additional beam through the second interferometer end mirror. The phase of the beam was modulated via a piezo-actuated mirror at a frequency of 5\,Hz and with a large motional amplitude of a few $\lambda$. This deep modulation led to fringe wrapping and frequency up-conversion, and produced a broadband disturbance over a bandwidth of about 200\,Hz. Such disturbances, that originate from sources at typically low frequencies but with large motional amplitudes, are a well known issue in GW-detectors \cite{Fiori2010,Ottaway2012,AdvDetBook,Martynov2015,AdvLIGOSens2016}. An electro-optic modulator (EOM) imprinted an additional phase modulation at 5.25 MHz on the back-scatter beam. As described above, both signals were frequency shifted to the MHz regime and demodulated before data acquisition (DAQ) to recover the audio-band signals. This way technical noise from the laser source that appeared for frequencies below $\approx 3\,\text{MHz}$ was avoided. Above that frequency our setup was limited by optical shot noise.

The interferometer was stabilized to a dark fringe and the output signal was split at a 50/50 beam splitter and detected with two balanced homodyne detectors (BHD1\&2). Each of these detectors used a strong external local oscillator field, whose phase with respect to the signal beam set the readout quadrature. One was measuring the phase quadrature and the other one the amplitude quadrature of the interferometer output field. 
The readout was enhanced with entangled, two-mode-squeezed states of light coming from the source described in \cite{SBM2013,10dBEPR2013}. We generated the two-mode-squeezing by overlapping two squeezed states at a 50/50 beam splitter with a relative phase shift of $90^{\circ}$. One of the output states we reflected at the dark port of the interferometer where it picked up the interferometer signal. Afterwards the two states recombined at the 50/50 beam splitter in the interferometer output path. By choosing the right phase relation between the two, the interference recovered the initial squeezed states. Both outputs carried the interferometer signal but one was squeezed in the amplitude and the other one in the phase quadrature of the interferometer signal. All degrees of freedom in our setup were electronically stabilized, except for the quadrature orientation of the orthogonally locked detectors with respect to the interferometer signal. This was adjusted by minimizing a marker peak in the spectrum of BHD1. The peak was a strong pure phase modulation at 1\,kHz and was generated with the same piezo actuated interferometer end mirror that also produced the GW-like signal. The data were acquired with a PCI-6259 card from National Instruments and processed in LabView.

\begin{figure}[t]
 \center 
  \includegraphics[width=\columnwidth]{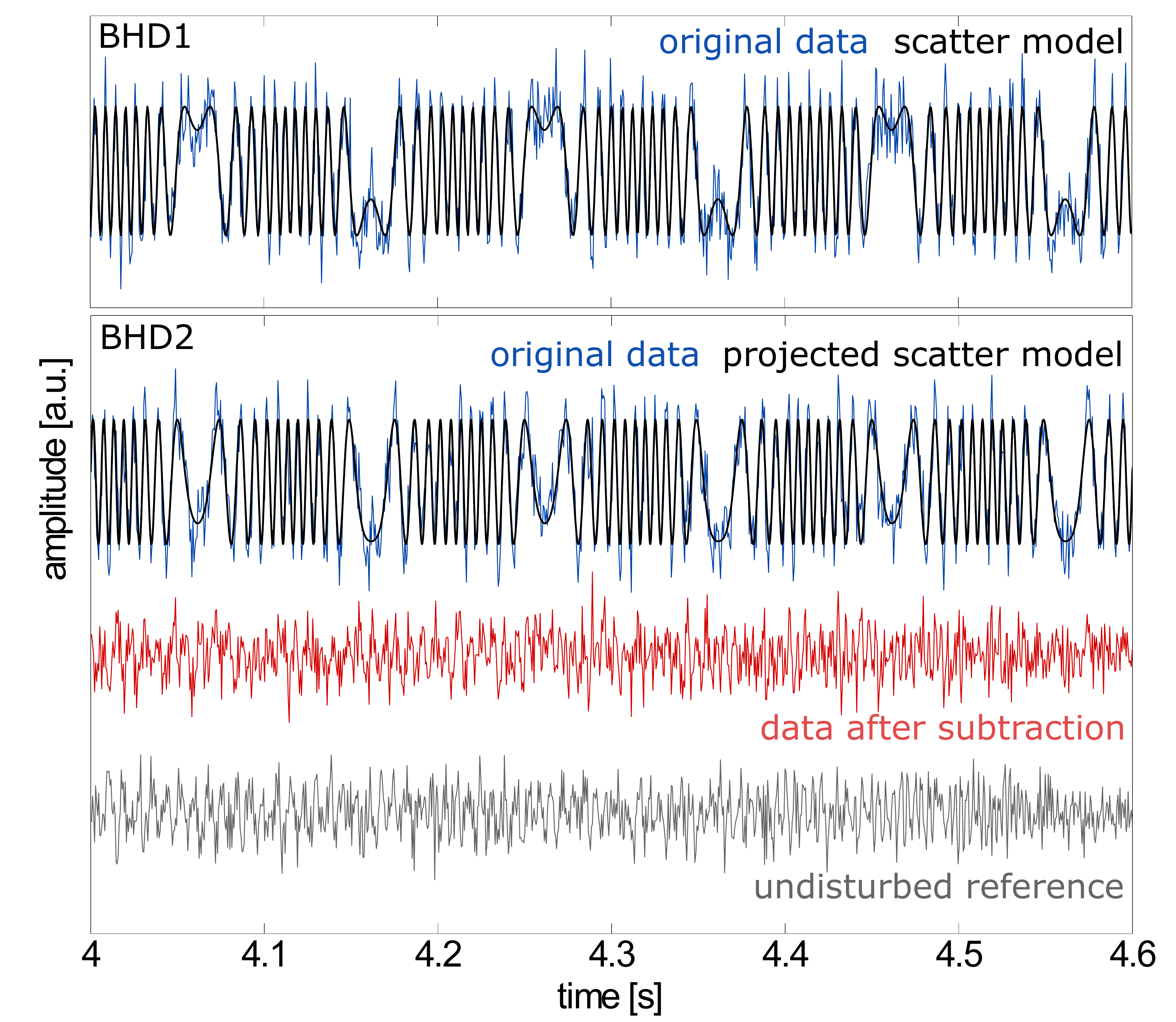}
  \caption{\textbf{Subtraction of the back-scatter disturbance in time domain.} The plots show the phase and amplitude quadrature measurement data of the respective detectors BHD1\&2 (blue), overlaid with the best fit of the back-scatter model (black). The amplitude quadrature data after subtraction of the disturbance are shown in red. For comparison, a reference measurement where the disturbance signal was blocked and only the GW-like signal was present is shown in gray. In both traces (red and gray), the injected chirp signal is just about discernible.}
  \label{Fig2-QDM-TimeDomainResults}
\end{figure}

\textbf{Time-domain data post-processing:} -- Fig.\,\ref{Fig2-QDM-TimeDomainResults} shows a segment of the measured time-domain data of both balanced homodyne detectors (blue lines). The data show a strong periodic signal which originates from the scatter disturbance and obscures the weaker GW-like signal.  
The post-processing of the data was done in Matlab, as described in \cite{ScM2015}. First, the parasitic interference was modeled and fitted to the phase quadrature data measured at BHD1. Our model of the scatter source was a sinusoidal motion with constant amplitude, frequency and phase including higher harmonics up to the 5th order. The latter turned out to be necessary to successfully describe the non-linear behavior of the piezo actuator that we used to generate the back-scatter disturbance. The time dependent phase shift of the back-scattered light can be modeled by 
\begin{equation}
\varphi(t)=\varphi_0+\tfrac{2\pi}{\lambda} \sum_{n=1}^{5}{m_n\,\sin(2\pi f t+\phi_n)^n} 
\label{eqPhi} 
\end{equation}
with the modulation frequency $f$, modulation depths $m_n$ and phases $\phi_n$ of the respective orders and an overall phase shift $\varphi_0$.
The projection of the resulting disturbance signal into the orthogonal quadratures at BHD1\&2 is given by 
\begin{align}
p_\text{sc}^\text{\tiny{BHD1}}(t)&= A \cos{\varphi(t)} \label{eqPProj} \\
x_\text{sc}^\text{\tiny{BHD2}}(t)&= A \sin{\varphi(t)} \label{eqXProj} 
\end{align}
where the amplitude $A$ depends on the intensity of the scattered light beam.
Fitting Eq.\,(\ref{eqPProj}) to the data of BHD1 provides all parameters that determine $\varphi(t)$. For the projection of the disturbance signal into the amplitude quadrature measurement via Eq.\,(\ref{eqXProj}), we allowed for slightly different amplitudes at the two detectors and an additional constant phase shift in Eq.\,(\ref{eqXProj}). This was done to compensate e.g. for an unbalanced splitting of the interferometer output and an imperfect quadrature orientation of the detectors.
The resulting fits are shown in black in Fig.\,\ref{Fig2-QDM-TimeDomainResults}. The amplitude quadrature data of BHD2 after subtraction of the modeled back-scatter disturbance is shown in red. The result already enables the recognition of a chirp, as expected for the injected GW-like signal. 
For comparison, a reference measurement is given in gray where the scattered light beam was blocked and only the GW-like signal was being injected.


\textbf{Frequency domain results:} -- 
\begin{figure}[t]
 \center 
  \includegraphics[width=\columnwidth]{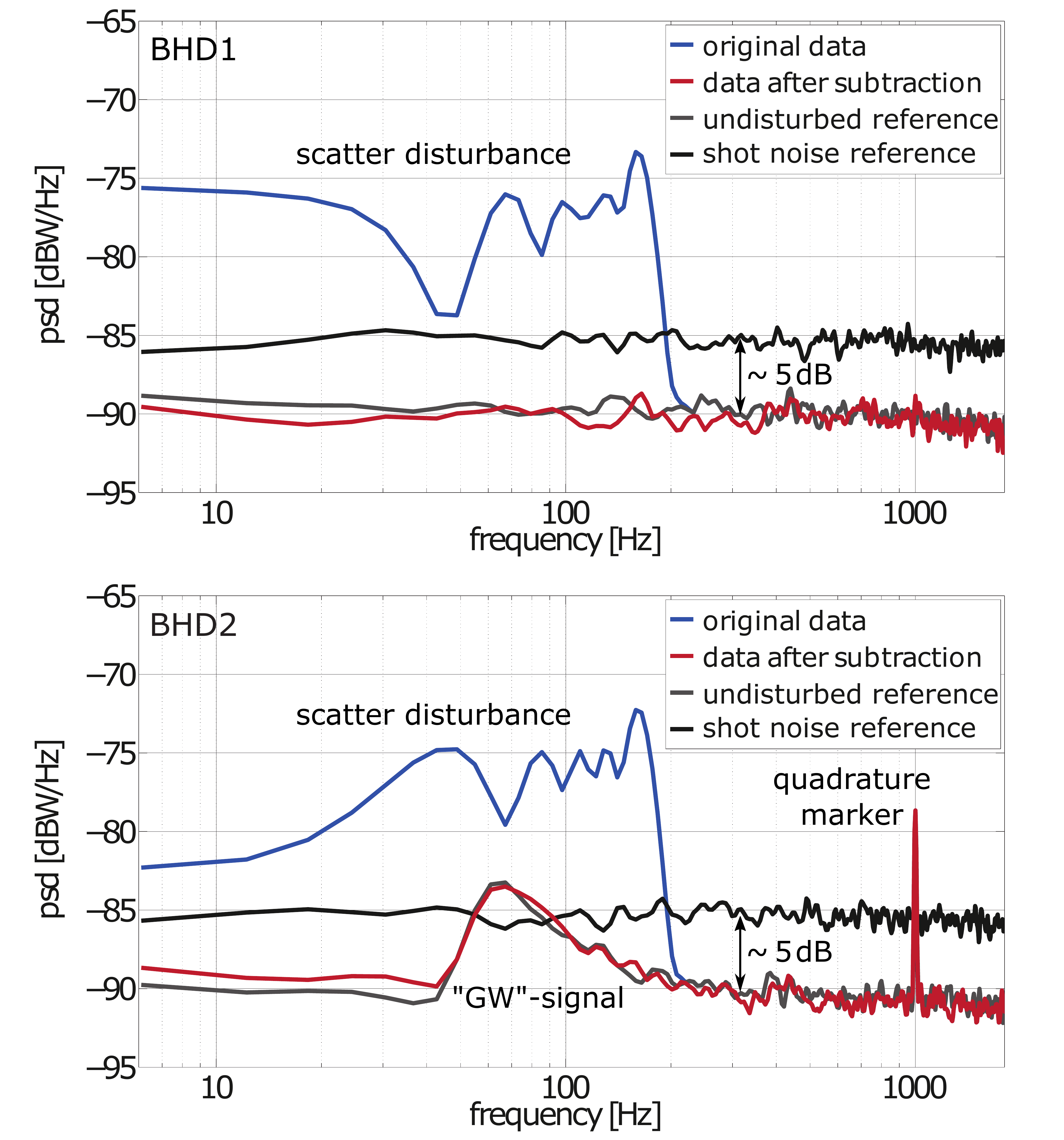}
  \caption{\textbf{Sensitivity improvement via QDM in frequency domain.} The plots show the averaged power spectral density (psd) of the respective measurements at the two readout detectors BHD1\&2. The x-axis corresponds to frequencies after demodulation with 5.25\,MHz. Depicted are the original data that contains the GW-like signal as well as the parasitic interference (blue), the same data after subtraction of the disturbance (red), a reference measurement with the disturbance blocked, containing only the GW-like signal (gray), and a shot noise measurement with the signal ports of the detectors blocked (black). After subtraction of the disturbance, a non-classical noise suppression of $\approx 5\,\text{dB}$ was achieved over the whole spectrum}
 \label{Fig3-QDM-FrequencyDomainResults}
\end{figure}%
Fig.\,\ref{Fig3-QDM-FrequencyDomainResults} shows the averaged power spectral density (psd) computed for the whole data sets of 5\,s length. The vacuum shot noise level was measured by blocking the signal ports at both detectors and is shown in black. The original measurement data, containing the GW-like signal and the back-scatter disturbance, are given in blue. Both, phase and amplitude quadrature measurement, showed sub-shot-noise sensitivity at (demodulated) frequencies above $\approx 200\,\text{Hz}$, corresponding to about 5\,dB of squeezing. Below that frequency, strong classical noise from the scattered light disturbance was limiting the sensitivity. The data after subtraction of the back-scatter disturbance is shown in red. Clearly, the broadband disturbance has vanished and the remaining `bump' visible in the data of BHD2 is in fact the injected GW-like signal. This becomes clear when we compare it to the reference measurement without the scatter disturbance which is shown in gray again. After subtraction of the classical disturbance, we achieved sub-shot-noise sensitivity at both detectors over the whole frequency range.

\begin{figure}[t]
 \center 
  \includegraphics[width=\columnwidth]{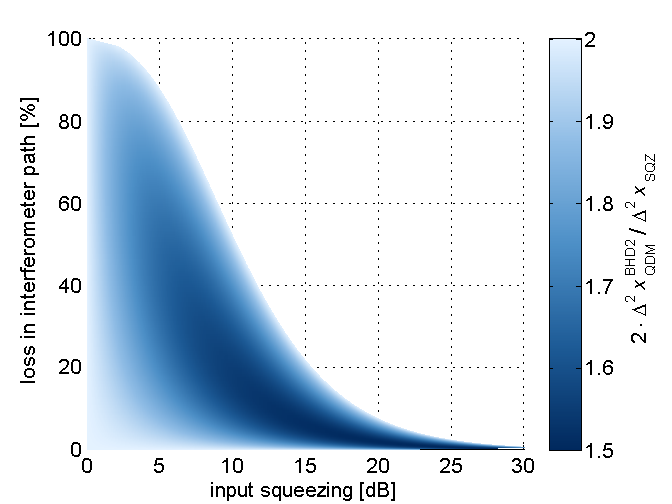}
  \caption{\textbf{Analysis of signal-to-(squeezed-)shot-noise ratio in QDM.} The (balanced) beam splitting necessary for QDM results in a reduction factor of 2 for zero dB squeezing (in power) compared with the single-quadrature squeezed readout \cite{Caves81,SqGEO2011}. In the presence of loss the factor is reduced down to a value of 1.5 if squeezing is applied. The plot assumes zero optical loss outside the interferometer. The uncolored region of the plot has values above 2. This regime is not shown since it is practically not relevant. In the presence of high optical loss, high squeezing values are only disadvantageous since they increase the influence of phase noise \cite{Suzuki2006,Franzen2006,Dwyer2013}. This is true regardless whether a single or two different readout quadratures are squeezed. Note that this analysis is only relevant for sideband frequencies with photon counting noise limited sensitivity whereas QDM targets frequencies being limited by excess noise.}
\label{Fig4-QDM-OpticalLoss}
\end{figure}

\textbf{Optical loss: } -- Optical loss on squeezed or two-mode-squeezed light reduces the observable squeezing factor. For a single pure squeezed field that is injected into a lossy interferometer \cite{Caves81} the squeezed variance of the amplitude quadrature at the photo detector is given by
\begin{equation}
\Delta^2 \hat{x}_\text{\tiny{SQZ}}(r,\eta_\text{ifo}) =  1-\eta_\text{ifo}+e^{-2r}\eta_\text{ifo} 
\label{eqSqzVar} 
\end{equation}
with $r$ the squeezing parameter and $\eta_\text{ifo}$ the full interferometer path efficiency. Here, the variance of a vacuum state is normalized to unity.
In QDM, the optical loss on the two entangled beams is asymmetric, with a higher loss on the beam that travels through the interferometer. 
Let us assume (i) negligible optical loss outside the interferometer, (ii) that the QDM scheme uses a \emph{balanced} splitter at the output port and (iii) that the additional readout is set to the orthogonal quadrature not containing any gravitational-wave signal. The variance of the squeezed amplitude quadrature at the relevant detector BHD2 is then given by
\begin{align}
\begin{split}
\Delta^2 \hat{x}_\text{\tiny{QDM}}^\text{\tiny{BHD2}}(r,\eta_\text{ifo}) &= \tfrac{1}{2}(1-\eta_\text{ifo}) + \tfrac{1}{4} e^{-2r} (\sqrt{\eta_\text{ifo}}+1)^2 \label{eqEntVar}\\
&+\tfrac{1}{4} e^{+2r} (\sqrt{\eta_\text{ifo}}-1)^2 \quad .
\end{split}
\end{align}
The fact that the beam outside the interferometer suffers much less optical loss can result in larger squeezing factors observed at the detectors when comparing with the injection of a single squeezed field as proposed by C.M. Caves and realized in GEO\,600 \cite{Caves81,SqGEO2011}. This benefit with respect to the single quadrature squeezed readout partly compensates for the fact that the QDM scheme requires splitting the interferometers output field into two (equal) parts, of which only one is used for an optimal GW signals readout. 
The actual reduction of (squeezed-)shot-noise-limited sensitivity accompanied with QDM is shown in Fig.~\ref{Fig4-QDM-OpticalLoss}. For zero dB squeezing, the signal-to-shot-noise ratio is deteriorated by 2, simply due to the balanced splitting of the output field in QDM. The corresponding color is light blue. However, for higher squeezing values and a certain range of optical loss, the signal-to-quantum-noise ratio reduces by smaller factors, down to 1.5. Achievable input squeezing factors of about 10\,dB \cite{Vahlbruch2008} combined with realistic optical loss values around 25\% for the transmission through a complex interferometer such as a GW detector, lie well within that range.


\textbf{Conclusion: } -- 
In conclusion, we report on the first proof-of-principle experiment in which the simultaneous readout of two orthogonal observables enables the improvement of measurement sensitivity from an excess noise limited regime to the sub-Poissonian (squeezed) regime. Our approach is based on Gaussian entanglement and targets noise sources in laser interferometers that produce excess noise in such a way that not only the quadrature angle of the signal is affected but also the orthogonal phase space projection. Back-scattered light (parasitic-interference) is of high relevance in current and future gravitational-wave detectors. The proof-of-principle reported here can be directly transferred to GW detectors once they include balanced homodyne detection for readout \cite{Fritschel2014,SBBDG2015a}. QDM requires a splitting of the interferometer's output field and reading out one beam with a quadrature angle that is not optimum for the actual (gravitational-wave) signal. We have shown that the accompanied loss in (squeezed-)shot-noise-limited sensitivity is less than the signal loss since the loss on the squeezing is actually reduced in realistic setups.
Our scheme is also of high relevance for identifying unknown noise sources in high-precision quantum metrology. In GW detectors it allows for discrimination between test mass displacement noise such as thermally driven test mass motion or radiation pressure noise and parasitic interferences. \\

\FloatBarrier

\begin{acknowledgments}
This work was partly supported by the Deutsche Forschungsgemeinschaft (Sonderforschungsbereich Transregio 7 and RTG1991) and by the International Max Planck Research School for Gravitational Wave Astronomy (IMPRS-GW). S. Steinlechner was supported by the Alexander von Humboldt foundation and the European Commission H2020-MSCA-IF-2014 actions, grant agreement number 658366.
\end{acknowledgments}




\begin{thebibliography}{99}
\newcommand{\enquote}[1]{``#1''}


\bibitem{AdvLIGO2015} The LIGO Scientific Collaboration, Advanced LIGO, \textit{Class. Quantum Gravity} \textbf{32}, 074001 (2015)

\bibitem{GEOHF2014} C. Affeldt \textit{et al.}, Advanced techniques in GEO 600, \textit{Class. Quantum Gravity} \textbf{31}, 224002 (2014)

\bibitem{AdvVirgo2015} F. Acernese \textit{et al.}, Advanced Virgo: a second-generation interferometric gravitational wave detector, \textit{Class. Quantum Gravity} \textbf{32}, 024001 (2015)

\bibitem{GW150914} The LIGO Scientific Collaboration and Virgo Collaboration, Observation of Gravitational Waves from a Binary Black Hole Merger, \textit{Phys. Rev. Lett.} \textbf{116}, 061102 (2016)

\bibitem{ScM2015} M. Meinders and R. Schnabel, Sensitivity improvement of a laser interferometer limited by inelastic back-scattering, employing dual readout, \textit{Class. Quantum Gravity} \textbf{32}, 195004 (2015)

\bibitem{Vahlbruch2007} H.~Vahlbruch, S.~Chelkowski, K.~Danzmann, R.~Schnabel, Quantum engineering of squeezed states for quantum communication and metrology, \textit{New Journal of Physics} \textbf{9}, 371 (2007)

\bibitem{Fiori2010} I. Fiori \textit{et al.}, Noise from scattered light in Virgo's second science run data. \textit{Class. Quantum Gravity} \textbf{27}, 194011 (2010)

\bibitem{Ottaway2012} D. J. Ottaway, P. Fritschel and S. J. Waldman, Impact of upconverted scattered light on advanced interferometric gravitational wave detectors, \textit{Opt. Express} \textbf{20}, Issue 8, 8329-8336 (2012)

\bibitem{AdvDetBook} M. Bassan (ed.), Advanced Interferometers and the Search for Gravitational Waves. Astrophysics and Space Science Library 404, DOI: 10.1007/978-3-319-03792-9\_1, Springer International Publishing Switzerland, 275-290 (2014)

\bibitem{Martynov2015} D. Martynov, Lock Acquisition and Sensitivity Analysis of Advanced LIGO Interferometers. \textit{PhD Thesis} California Institute of Technology, 115-122 (2015)

\bibitem{AdvLIGOSens2016} The LIGO Scientific Collaboration, The Sensitivity of the Advanced LIGO Detectors at the Beginning of Gravitational Wave Astronomy, arXiv:1604.00439v1 (2016)

\bibitem{Schnabel2010} R.~Schnabel, N.~Mavalvala, D.~E. McClelland, P.~K. Lam, Quantum metrology for gravitational wave astronomy, \textit{Nature communications} \textbf{1}, 121 (2010)
  
\bibitem{SqGEO2011} The LIGO Scientific Collaboration, A gravitational wave observatory operating beyond the quantum shot-noise limit, \textit{Nat. Phys.} \textbf{7},  962-965 (2011)

\bibitem{ltSqGEO2013} H. Grote, K. Danzmann, K. L. Dooley, R. Schnabel, J. Slutsky and H. Vahlbruch, First Long-Term Application of Squeezed States of Light in a Gravitational-Wave Observatory, \textit{Phys. Rev. Lett.} \textbf{110}, 181101 (2013)

\bibitem{SqLIGO2013} The LIGO Scientific Collaboration, Enhanced sensitivity of the LIGO gravitational wave detector by using squeezed states of light, \textit{Nat. Phot.} \textbf{7}, 613-619 (2013)

\bibitem{LIGOWhitePaper2015} The LIGO Scientific Collaboration, Instrument Science White Paper, https://dcc.ligo.org/LIGO-T1400316/public, 13-14 (2015)

\bibitem{SBM2013}  S. Steinlechner, J. Bauchrowitz,  M. Meinders, H. M\"uller-Ebhardt, K. Danzmann and R. Schnabel, Quantum-dense metrology, \textit{Nat. Phot.} \textbf{7},  626-630 (2013)

\bibitem{gwSound} Simulated sound of binary neutron stars from the research group of  Professor Scott A. Hughes at MIT, http://gmunu.mit.edu/sounds/comparable\_sounds/\\comparable\_sounds.html

\bibitem{10dBEPR2013} S. Steinlechner, J. Bauchrowitz, T. Eberle and R. Schnabel, Strong Einstein-Podolsky-Rosen steering with unconditional entangled states, \textit{Phys. Rev. A} \textbf{87}, 022104 (2013).

\bibitem{Caves81} C. M. Caves, Quantum-mechanical noise in an interferometer, \textit{Phys. Rev. D} \textbf{23}, 1693 (1981)

\bibitem{Vahlbruch2008} H. Vahlbruch, M. Mehmet, S. Chelkowski, B. Hage, A. Franzen, N. Lastzka, S. Go{\ss}ler, K. Danzmann and R. Schnabel, Observation of Squeezed Light with 10-dB Quantum-Noise Reduction, \textit{Phys. Rev. Lett.} \textbf{100}, 033602 (2008).

\bibitem{Suzuki2006} S. Suzuki, H. Yonezawa, F. Kannari, M. Sasaki and A. Furusawa, 7dB quadrature squeezing at 860 nm with periodically poled KTiOPO, \textit{Applied Physics Letters} \textbf{89}, 061116 (2006).
\bibitem{Franzen2006} A. Franzen, B. Hage, J. DiGuglielmo, J. Fiur\'a\v{s}ek and R. Schnabel, Experimental Demonstration of Continuous Variable Purification of Squeezed States, \textit{Phys. Rev. Lett.} \textbf{97}, 150505 (2006).
\bibitem{Dwyer2013} S. Dwyer \textit{et al.}, Squeezed quadrature fluctuations in a gravitational wave detector using squeezed light, \textit{Opt. Express} \textbf{21}, 19047-60 (2013).
 
\bibitem{Fritschel2014} P. Fritschel, M. Evans and V. Frolov, Balanced homodyne readout for quantum limited gravitational wave detectors, \textit{Opt. Express} \textbf{22}, 4224-4234 (2014).
\bibitem{SBBDG2015a} S. Steinlechner \textit{et al.}, Local-oscillator noise coupling in balanced homodyne readout for advanced gravitational wave detectors, \textit{Phys. Rev. D} \textbf{92}, 072009 (2015).

\end{thebibliography}
\end{document}